# Exploration of Graph Computing in Power System State Estimation

Chen Yuan[a], Yuqi Zhou[a,b], Guofang Zhang[c], Guangyi Liu[a], Renchang Dai[a], Xi Chen[a], Zhiwei Wang[a]
[a] GEIRI North America, San Jose, CA, USA
[b] Texas A&M University, College Station, TX, USA
[c] State Grid Sichuan Electric Power Company, Chengdu, Sichuan, China
guangyi.liu@geirina.net

*Abstract*—With the increased complexity of power systems due to the integration of smart grid technologies and renewable energy resources, more frequent changes have been introduced to system status, and the traditional serial mode of state estimation algorithm cannot well meet the restrict time-constrained requirement for the future dynamic power grid, even with advanced computer hardware. To guarantee the grid's reliability and minimize the impacts caused by system status fluctuations, a fast, even SCADA-rate, state estimator is urgently needed. In this paper, a graph based power system modeling is firstly explored and a graph computing based state estimation is proposed to speed up its performance. The power system is represented by a graph, which is a collection of vertices and edges, and the measurements are attributes of vertices and edges. Each vertex can independently implement local computation, like formulations of the node-based H matrix, gain matrix and right-hand-side (RHS) vector, only with the information on its connected edges and neighboring vertices. Then, by taking advantages of graph database, these node-based data are conveniently collected and stored in the compressed sparse row (CSR) format avoiding the complexity and heaviness introduced by the sparse matrices. With communications and synchronization, centralized computation of solving the weighted least square (WLS) state estimation is completed with hierarchical parallel computing. The proposed strategy is implemented on a graph database platform. The testing results of IEEE 14-bus, IEEE 118-bus systems and a provincial system in China verify the accuracy and high-performance of the proposed methodology.

*Index Terms*--Graph database, sparse matrices, state estimation, weighted least square (WLS).

## I. INTRODUCTION

Power system state estimation refers to obtaining the voltage phasors of all system buses, including voltage magnitudes and phase angles, at a given moment. This is generally performed by making observations of the power flows through the network, then performing an inference to estimate values of the underlying voltage phasors [1]. The outputs of system state estimation are critical for monitoring system status and updating the real-time model for subsequent applications in energy management system (EMS), like contingency analysis, optimal power flow, economic dispatch, unit commitment, etc. In today's practice, state estimation runs every 1-5 minutes for large power systems, leading to a 1-5 minutes delay between the present power system and the estimated one. If a severe event happens, the estimated system status may have a large difference from the real-time system status and then it is very difficult for system operators to timely identify the problem and secure the system. Moreover, as the increased complexity of power systems due to the penetrations of renewable energy resources [2], intelligent control devices [3] and smart meters [4], more frequent changes have been introduced to systems status. Department of Energy (DOE) released computational needs for next generation electric grid in [5], the future direction is to achieve real-time state estimation within one SCADA measurement cycle. An accurate and faster state estimation will allow operators to keep monitoring system status changes and then implement subsequent EMS applications to improve the system's reliability and robustness. High-performance computing technology makes the faster processing of state estimation possible. Parallel computing is one effective procedure, especially for large interconnected power systems.

The idea of multiprocessor state estimation was proposed in [6]. Reference [7] discusses the possibility of distributed state estimation. Parallel state estimation based on fast decoupled method was illustrated in [8]. A block Jacobian method was used in [9] to implement distributed state estimation. A decentralized robust state estimator was proposed in [10]. The matrix inversion lemma was used for parallel static state estimation in [11] and the block-partitioning algorithm was also proposed. The algorithm gave an exact solution, but the partitioning will have a vital impact on the success of the parallel algorithm. The work conducted by Pacific Northwest National Laboratory (PNNL) applies high performance computing and adapted algorithm to speed up state estimation program [12]. Its focus is on how to efficiently solve the state estimation equation. A matrix-splitting strategy for Gauss-Newton iteration was presented in [13], in which neighbor areas communicate with each other during each iteration. A computationally efficient linear model-based estimator was proposed in [14] to serve as an alternative for Gauss-Newton approach. High-performance computing (HPC) with GPUs

This work is supported by the State Grid Corporation technology project SGSHXT00JFJS1700138.

were used in [15], in which a parallel relaxation-based joint state estimation algorithm was proposed for dynamic state estimation.

In this paper, the graph computing technology is explored and applied to power system state estimation to increase the computing efficiency and realize fast, even SCADA-rate, state estimation for large-scale systems. The proposed graph computing methodology takes advantages of graph database to efficiently formulate H matrix, gain matrix and right-hand-side (RHS) vector. According to system topology analysis and decomposition of the state estimation problem, the node-based local computation only needs information from its neighbors within 2 steps, which are easily traversed in 1-2 graph operations. Besides, as an equivalent natural format of system topology, compressed sparse row (CSR) is employed to store computation matrices for further improving calculation efficiency and relieving the burden introduced by matrix sparsity. The gain matrix factorization is implemented by hierarchical parallel computing.

The paper is organized as follows. In Section II, graph theory is briefly introduced, and graph computing is explored for power system applications. The graph computing based WLS state estimation is presented in Section III, and Section IV verifies the proposed methodology's accuracy and high-performance with testing results. The conclusion is given in Section V.

## II. GRAPH COMPUTING

### A. Graph Theory and Graph Database

Graph theory is the study of graphs, which are used to model relations between objects. A graph is a collection of vertices, representing objects in a system, and edges, standing for relations between objects. It is represented as $G = (V, E)$, in which $V$ indicates a set of vertices in the graph, $G$, and the set of edges in the graph is represented as $E$, expressing how these vertices relate to each other. Each edge is denoted by $e = (i, j)$ in $E$, where we refer to $i$ and $j$ in $V$ as head and tail of the edge $e$, respectively.

Graph database uses graph structures for semantic queries with vertices, edges and attributes to represent and store data in vertices and edges. Such database allows data in the store to be linked together directly, and retrieved with one operation.

### B. Graph Computing Applications in Power Systems

Graphs are useful in understanding and processing diverse data and they have already been applied to model relations and processes in physical, biological, social and information systems. This paper proposes to use the graph database to naturally represent power systems and apply graph computing to energy management systems in power grids. Generally, a traditional power system is comprised of power generation, power delivery, power distribution, power conversion and power consumption. The five components are categorized into two types: (1) bus-attached, and (2) line-attached. For example, generators and loads, are connected to buses, regarded as power injections at buses. So, they are taken as bus-attached components. Power delivery and distribution are based on power transmission and distribution lines, acting as line-attached parts. Similarly, for power conversion part, since it is employed to make voltage conversion, AC-DC conversion is also thought as a type of voltage conversion, between its terminals, it is assigned as a line-attached component.

*1) Node-Based Parallel Computing*: In graph computing, node-based parallel computing means computation at each node is independent from each other, meaning such local computations are performed simultaneously [16]. Fig. 1 depicts relations between graph computation and matrix computation and the Bulk Synchronous Parallel (BSP) based graph computing strategy. It shows that, in a graph, the connections between nodes represent non-zeros in off-diagonal elements. Regarding each zero in the matrix A, it means no relations or impacts between its two corresponding nodes, so there is no connection displayed between these two nodes in the graph topology, neither. For the node-based parallel computing, take the admittance matrix as an example, the off-diagonal element in the admittance matrix is locally computed based on the attribute of the corresponding edge, and the diagonal element is obtained by processing the attributes on the node and its connected edges. Therefore, all elements in the admittance matrix are computed within one node-based graph operation. Other examples of node-based parallel computation in power systems are node active power and reactive power injection calculation, node variables mismatch calculation, convergence check at each iteration and post-convergence branch active power and reactive power flow calculation. These calculations on each node are independent to other nodes.

*2) Hierarchical Parallel Computing:* Hierarchical parallel computing performs computation on nodes at the same level in parallel. The level next to it is performed after. The hierarchical parallel computing is applied to matrix factorization. To factorize matrix using Cholesky elimination algorithm, three steps are involved for hierarchical parallel computing: 1) determining fill-ins, 2) forming elimination tree, and 3)

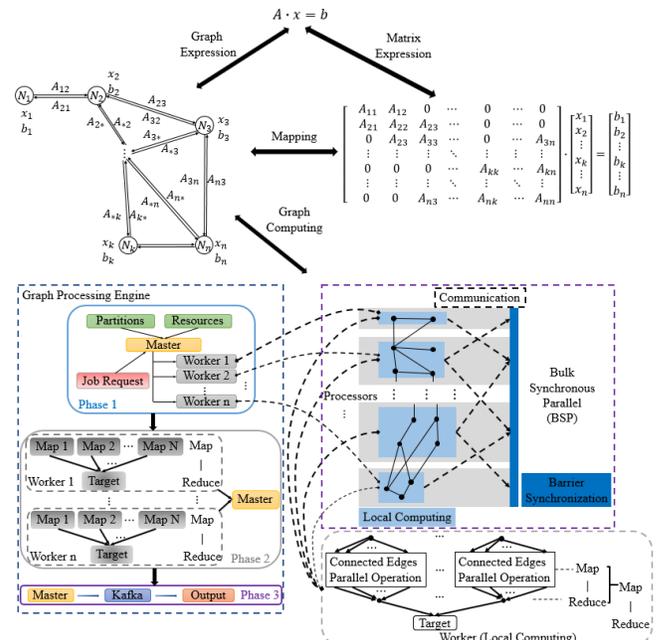

Figure 1. Mapping between graph computation and matrix computation

partitioning elimination tree for hierarchical parallel computing.

## III. Graph Computing based WLS State Estimation

### A. WLS State Estimation

The weighted least square (WLS) algorithm is the most widely used methodology in existing state estimators to minimize the weighted sum of the square of residuals between the actual measurements and estimations. The measurement model in power system state estimation is presented below [1].

$$z = h(x) + e \tag{1}$$

where $z$ is the measurements vector, $x$ is the system state vector, including magnitudes and phase angles of bus voltages, $h(x)$ is the nonlinear measurement function vector and $e$ is a vector of measurement errors, being considered as Gaussian noise with zero means. The state variables considered here are bus voltage magnitudes and phase angles while the measurements are bus voltage magnitudes, bus power injections and power flows. For a n-bus power system, assuming it has m lines, then there are $3n+4m$ measurements, i.e., $n$ bus voltage magnitudes, $2n$ power injects at buses and $4m$ power flows. Regarding the state variables, since the phase angle at the slack bus is considered as a reference value, there are $2n-1$ states to be estimated. Nonlinear WLS state estimation is then formulated as:

minimize: $J(x) = [z - h(x)]^T \cdot R^{-1} \cdot [z - h(x)]$

subject to: $z = h(x) + e$

The following equation is obtained at the minimum:

$$g(x) = \frac{\partial J(x)}{\partial x} = -H^T(x) \cdot R^{-1} \cdot [z - h(x)] = 0 \tag{2}$$

where $g(x)$ is the matrix of the gradient of the objective function $J(x)$, $H(x) = \frac{\partial h(x)}{\partial x}$ is the Jacobian matrix of $h(x)$ and $R^{-1}$ is the weight matrix. Substituting the first-order Taylor's expansion of $g(x)$ in (2), the following equation (3) is iteratively solved to find the solution that minimizes $J(x)$.

$$\Delta x = [G(x^k)]^{-1} \cdot H^T(x^k) \cdot R^{-1} \cdot (z - h(x^k))$$
$$\rightarrow G(x^k) \cdot \Delta x = H^T(x^k) \cdot R^{-1} \cdot (z - h(x^k)) \tag{3}$$

where,

$$g(x^{k+1}) = g(x^k) + G(x^k) \cdot (x^{k+1} - x^k) = 0 \tag{4}$$

$$G(x^k) = \frac{\partial g(x^k)}{\partial x} = H^T(x^k) \cdot R^{-1} \cdot H(x^k) \tag{5}$$

$$x^{k+1} = x^k + \Delta x \tag{6}$$

### B. Graph Computing for WLS State Estimation

In an n-bus system, the system measurements are divided into $n$ parts and assigned to each node. The measurement model for each partition $i$ is written as:

$$z_i = h_i(x) + e_i, \; i = 1, \cdots, n \tag{7}$$

$$z_i = [V_i \; P_i \; Q_i \; P_{ij} \; Q_{ij}]^T, j \neq i \tag{8}$$

Based on the above partition, the relation between system-level measurement model and node-based one is shown in (9), and the weight matrix is block-partitioned as displayed in (10).

$$z = [z_1^T \; z_2^T \; \cdots \; z_n^T]^T \tag{9}$$

$$R = \begin{bmatrix} [R_1^{-1}] & 0 & \cdots & 0 \\ 0 & [R_2^{-1}] & \ddots & \vdots \\ \vdots & \ddots & \ddots & 0 \\ 0 & \cdots & 0 & [R_n^{-1}] \end{bmatrix} \tag{10}$$

where $R_i^{-1}$ is the weight matrix for measurements associated with node $i$ and its dimension is $m_i \times m_i$, $m_i$ is the number of measurements related to node $i$, including $V_i$, $P_i$, $Q_i$, $P_{ij}$, $Q_{ij}$. The equations of node-based residual are as follows:

$$r_i = z_i - h_i(x) \tag{11}$$

$$r = [r_1^T \; r_2^T \; \cdots \; r_n^T]^T \tag{12}$$

The corresponding $H$ matrix and $G$ matrix of node $i$ is given by:

$$H_i(x) = \frac{\partial h_i(x)}{\partial x} \tag{13}$$

$$G_i(x) = H_i^T(x) R_i^{-1} H_i(x) \tag{14}$$

where $G_i(x)$ is the gain matrix for node $i$ and it has the same dimension as the system gain matrix $G(x)$, which is proved as follows.

$$H(x) = \frac{\partial h(x)}{\partial x}$$
$$= \frac{\partial [h_1(x)^T \; h_2(x)^T \; \cdots \; h_n(x)^T]^T}{\partial x}$$
$$= \left[\left(\frac{\partial h_1(x)}{\partial x}\right)^T \; \left(\frac{\partial h_2(x)}{\partial x}\right)^T \; \cdots \; \left(\frac{\partial h_n(x)}{\partial x}\right)^T\right]^T$$
$$= [H_1(x)^T \; H_2(x)^T \; \cdots \; H_n(x)^T]^T \tag{15}$$

$$G(x) = H^T(x) \cdot R^{-1} \cdot H(x)$$
$$= [H_1(x)^T \; H_2(x) \; \cdots \; H_n(x)^T] \cdot R^{-1}$$
$$\cdot [H_1(x)^T \; H_2(x)^T \; \cdots \; H_n(x)^T]^T$$
$$= \sum_{i=1}^n H_i^T(x) \cdot R_i^{-1} \cdot H_i(x)$$
$$= \sum_{i=1}^n G_i(x) \tag{16}$$

The $h(x)$ and $H$ matrix in (15) are reordered with the node-based rearrangement of system measurements and the sequence difference of $H$ matrix does not impact system gain matrix, since the essence of gain matrix formulation is the multiplication of column vectors of $H$ matrix. Therefore, the system H matrix and gain matrix are formulated in parallel based on independent local computation at each node. Combining with the BSP based graph computing shown in Fig. 1, the proposed node-based graph computing strategy is implemented and applied to state estimation problem.

Furthermore, the RHS vector of (3) is executed in node-based parallel as well.

$$H^T(x^k) R^{-1}(z - h(x^k)) = \sum_{i=1}^n H_i^T(x^k) R_i^{-1} r_i \tag{17}$$

After the formulation of gain matrix and RHS vector, Hierarchical parallel computing based LU decomposition is employed to achieve triangular matrices, L and U, of gain matrix. Then forward and backward substitution is used to get $\Delta x$ and then update $x$.

The equations (6) - (17) indicate the feasibility of node-based graph computing in power system state estimation. The desire that we derive these equations is to realize independent computation of the node-based H matrix, $H_i(x)$, gain matrix, $G_i(x)$, and RHS vector, $H_i^T(x^k)R_i^{-1}r_i$, in parallel. Then these node-based matrices are easily aggregated to form the system-level gain matrix, $G(x)$, and RHS vector, $H^T(x^k)R^{-1}(z - h(x^k))$, for centralized state estimation.

Without further dealing with the matrix sparsity, the proposed node-based graph computing method already partitions the computation into $n$ parts so that the computing time complexity is largely reduced. This is because the computational complexity for calculating system gain matrix and node-based gain matrix are $O(M^2 n^2)$ and $O(M_i^2 n_i^2)$ respectively, where $n$ is the number of nodes, $M$ is the total measurements of the system, $M_i$ is the size of $z_i$ and $n_i$ is the number of selected nodes, including node $i$ and its neighbors. It can be observed that the latter is much smaller than the former one. With node-based parallel computing, the complexity of the whole system gain matrix is also significantly reduced.

Even though the computation efficiency is much lifted with node-based partitions and graph computing, these node-based H matrix, $H_i(x)$, and gain matrix, $G_i(x)$, are still very sparse. In other words, the sparsity of H matrix and gain matrix haven't been changed with the node-based partitions, meaning there is a necessity to relieve the burden of sparse matrices storage and then further improve state estimation performance. The next section will discuss how to use graph database to deal with the sparse matrices and how efficient graph computing is for power system state estimation.

### C. System Topology Analysis for Dense Gain Matrix Formulation Using Node-Based Graph Computing

As presented in equations (18) and (19), power injections, $P_i$ and $Q_i$, depend on the status of bus $i$ itself, its 1-step neighboring buses status, and the line configurations between bus $i$ and its 1-step neighbors. Similarly, power flows, $P_{ij}$ and $Q_{ij}$, depend on the configuration of the line, $L_{ij}$, and its terminal buses status, as displayed in (20) and (21).

$$P_i = V_i \sum V_j(G_{ij} \cos \theta_{ij} + B_{ij} \sin \theta_{ij}) \quad (18)$$

$$Q_i = V_i \sum V_j(G_{ij} \sin \theta_{ij} - B_{ij} \cos \theta_{ij}) \quad (19)$$

$$P_{ij} = V_i^2(g_{si} + g_{ij}) - V_i V_j(g_{ij} \cos \theta_{ij} + b_{ij} \sin \theta_{ij}) \quad (20)$$

$$Q_{ij} = -V_i^2(b_{si} + b_{ij}) - V_i V_j(g_{ij} \sin \theta_{ij} - b_{ij} \cos \theta_{ij}) \quad (21)$$

In power systems, sensitivity of real/reactive power equations to changes in the magnitude/phase angle of bus voltages is very low, especially for high voltage transmission system, whose X/R is large enough to ignore the line resistance. Furthermore, based on the observation, the elements of the gain matrix do not significantly change between flat start and the converged solution. Therefore, the fast-decoupled formulation of the state estimation problem is derived with the approximated but constant H matrix and gain matrix [1]. Based on equations (18) – (21) and fast-decoupled method, only the derivatives of bus voltages, $\frac{\partial V_i}{\partial V_i} = 1$, bus power injections, $\frac{\partial P_i}{\partial \theta_i}$ and $\frac{\partial Q_i}{\partial V_i}$, and power flows, $\frac{\partial P_{ij}}{\partial \theta_i}$, $\frac{\partial P_{ij}}{\partial \theta_j}$, $\frac{\partial Q_{ij}}{\partial V_i}$ and $\frac{\partial Q_{ij}}{\partial V_j}$, are non-zero constants.

$$H(x) = \begin{bmatrix} H_P & 0 \\ 0 & H_Q \end{bmatrix}, \quad G(x) = \begin{bmatrix} G_P & 0 \\ 0 & G_Q \end{bmatrix}$$

In equations (12) and (13), $H_i(x)$ only has nonzero elements in the rows and columns related to node $i$ and its 1-step neighbors, and $G_i(x)$ only has non-zero elements in the columns and rows of node $i$, its 1-step neighbors, and its 2-step neighbors. With the knowledge of graph computing, this paper continues to develop the node-based H matrix, $H_i(x)$, and gain matrix, $G_i(x)$, into dense matrices by eliminating zero elements and only storing non-zero elements. Replacing conventional sparse matrices with node-based dense matrices greatly saves storage space and improves data read/write speed. Besides, these node-based matrices are developed by fetching local information in parallel. Based on the characteristics of graph database, in order to further facilitate the built-up process of the system gain matrix, $G(x)$, each row vector of the obtained node-based gain matrix, $G_i(x)$, is distributed to the node, which is correspondingly mapped to the row vector. The row vectors at each node are summed to form a new one using the function of accumulator. In other words, the overlapped elements of each node's row vectors are summed up. Once the distribution of row vectors is completed, the system gain matrix is then easily formed by placing each row vector to the corresponding row in the system gain matrix, $G(x)$.

Take the IEEE 5-bus system as an example, as shown in Fig. 2, the structure and formation of gain matrix is further analyzed and elaborated as follows based on graph computing. For bus 1, its 1-step neighbors are bus 2 and bus 3. Bus 1's fast decoupled H matrix, $H_1$, is composed of two off-diagonal zero blocks and two non-zero diagonal parts, $H_{1P}$ and $H_{1Q}$, which are blocks respectively related to active power and reactive power.

$$H_1 = \begin{bmatrix} H_{1P} & 0 \\ 0 & H_{1Q} \end{bmatrix}$$

$$H_{1P} = \begin{bmatrix} \frac{\partial h_{1P}}{\partial \theta_1} & \frac{\partial h_{1P}}{\partial \theta_2} & \frac{\partial h_{1P}}{\partial \theta_3} \end{bmatrix}$$

$$H_{1Q} = \begin{bmatrix} \frac{\partial h_{1Q}}{\partial V_1} & \frac{\partial h_{1Q}}{\partial V_2} & \frac{\partial h_{1Q}}{\partial V_3} \end{bmatrix}$$

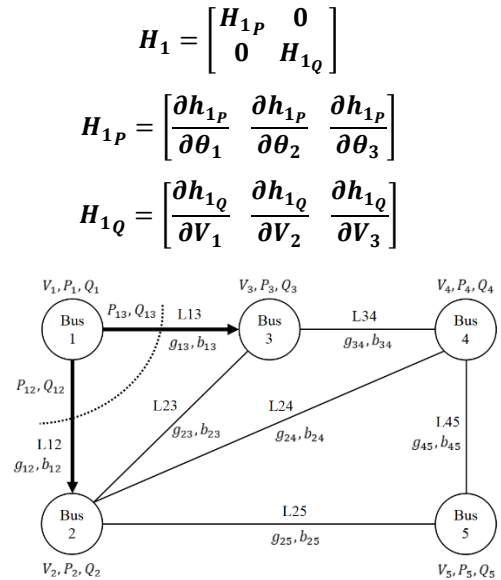

Figure 2. Graph based IEEE 5-bus system

$$G_1 = H_1^T R^{-1} H_1 = \begin{bmatrix} H_{1_P}^T R_P^{-1} H_{1_P} & 0 \\ 0 & H_{1_Q}^T R_Q^{-1} H_{1_Q} \end{bmatrix}$$

$$= \begin{bmatrix} G_{1_P} & 0 \\ 0 & G_{1_Q} \end{bmatrix}$$

where $h_{1_P} = [P_1 \ P_{12} \ P_{13}]^T$, $h_{1_Q} = [V_1 \ Q_1 \ Q_{12} \ Q_{13}]^T$, and $H_{1_P}$ is a $3 \times 3$ matrix, $H_{1_Q}$ is a $4 \times 3$ matrix, and $G_{1_P}$ and $G_{1_Q}$ are both $3 \times 3$ matrices. Similarly, $G_{2_P}, \cdots, G_{3_P}$ and $G_{2_Q}, \cdots, G_{3_Q}$ are obtained. Their sizes are $(m+1) \times (m+1)$, where $m$ is the number of the 1-step neighbors. In addition, take $G_{1_P}$ as an example, the element $G_{ij_P}^1$ denotes the $G_{1_P}$ contribution to the element in the row $i$ and column $j$ of the system-level decoupled gain matrix, $G_P$. Similarly, $G_{2_P}$ and $G_{3_P}$ have non-zero elements in row 1 of $G_P$. Therefore, the first row of $G_P$, $G_P(1)$, has non-zero elements in columns 1, 2, 3, 4, and 5. In other words, $G_P(1)$ has non-zero elements contributed by itself, its 1-step neighbor(s) and 2-step neighbor(s).

$$G_{1_P} = \begin{bmatrix} G_{11_P}^1 & G_{12_P}^1 & G_{13_P}^1 \\ G_{21_P}^1 & G_{22_P}^1 & G_{23_P}^1 \\ G_{31_P}^1 & G_{32_P}^1 & G_{33_P}^1 \end{bmatrix} \begin{matrix} 1 \\ 2 \\ 3 \end{matrix}$$

## IV. CASE STUDY

To explore the feasibility of graph computing in power system state estimation, the proposed graph computing methodology is implemented in a Linux server. The test environment is listed in Table I.

The performance of the proposed method is demonstrated in Table II, including testing in IEEE 14-bus system, IEEE 118-bus system and one provincial system in China, FJ-1425 case. As shown in Table II, the estimation results are very close to the system measurements with negligible mean squared errors. On the other hand, the speed of the graph computing based state estimation is fast. The time spent on residual update, equation solving, and RHS vector formation takes over 80% of the total time consumption.

## V. CONCLUSION

The exploration of power system modeling using graph database was presented in this paper and the graph computing based power system state estimation has been proposed to speed up its performance. The testing results in IEEE standard systems and a practical system verified the proposed method's accuracy and high-performance. In the future work, the graph based state estimation problem will be further analyzed to reduce the time spent on residual update, RHS Vector formulation and equation solving.

TABLE I. TEST ENVIRONMENT

| Hardware Environment | |
|---|---|
| CPU | 2 CPUs × 6 Cores × 2 Threads @ 2.10 GHz |
| Memory | 64 GB |
| **Software Environment** | |
| Operating System | CentOS 6.8 |
| Graph Database | TigerGraph v0.8.1 |

TABLE II. PERFORMANCE EVALUATION OF IEEE STANDARD

| | | IEEE 14-Bus System | IEEE 118-Bus System | FJ-1425 System |
|---|---|---|---|---|
| Flat Start | | YES | YES | YES |
| Number of Iterations | | 5 | 5 | 7 |
| System Measurements Mean Squared Error | Voltage Magnitude (p.u.) | 0 | 6.78E-14 | 1.42E-12 |
| | Voltage Angle (degree) | 7.03E-8 | 3.75E-9 | 5.97E-10 |
| Computation Time (ms) | Gain Matrix Formulation | 1.15 | 3.24 | 14.41 |
| | Gain Matrix Factorization | 0.41 | 1.50 | 9.11 |
| | Residual Update + Forward and Backward Substitution (One Iteration) | 1.45 | 3.74 | 10.61 |
| | RHS Vector (One Iteration) | 0.74 | 1.90 | 6.74 |
| | Total | 12.63 | 33.69 | 146.37 |